\documentclass[aps,a4paper,twocolumn,floatfix,showkeys,
groupaddress]{revtex4}

\usepackage{bm,amsmath,amssymb,amsfonts}

\usepackage[dvips]{graphicx}
\usepackage[colorlinks=true,citecolor=red,urlcolor=red,filecolor=blue]{hyperref}
\usepackage{color}
\definecolor{mycolor}{rgb}{0.80,0.10,0.20}
\begin{document}

\title{\textcolor{mycolor}
{Flux modulated flat band engineering in square-kagom\'{e} ladder network}}  

\author{Atanu Nandy}
\email{atanunandy1989@gmail.com}
\affiliation{Department of Physics, Kulti College, Kulti, Paschim Bardhaman,
West Bengal-713 343, India}

\author{Amrita Mukherjee}
\email{amritaphy92@gmail.com}
\affiliation{Department of Physics, University of Kalyani, Kalyani,
West Bengal-741 235, India}

\begin{abstract}
The origin of non-dispersive flat band modes for a quasi-one dimensional square-kagom\'{e}
 ladder network is
explored analytically by virtue of the real space renormalization group (RSRG) technique. 
A section of the eigenstates is non-diffusive i.e., localized within a cluster
of sub-lattice sites partly by the destructive type of quantum interference and partly
 by the physical boundary created
by the site with zero wave function amplitude. By making the amplitude vanish at the selective
 sites it becomes possible to confine the
incoming excitation within the \textit{trapping cell} leading to the formation of compact localized states. The effective mass
of the particle becomes infinitely large corresponding to those self-localized modes and 
hence the mobility of the
wave train becomes vanishingly small. This quenched kinetic energy leads to a momentum independent contribution
to a dispersion curve. The present analysis is corroborated by numerical calculation of spectral landscape and the
corresponding dispersion profile. 
The application of uniform magnetic flux may lead to a comprehensive engineering of the position as well
as the curvature of the band.
Also, one-to-one mapping between electronic case and photonic case within the
tight-binding framework helps us to study the photonic localization in an analogous single mode wave guide system.
The concept of slow light eventually introduces the possibility of spatial compression of light energy.
\end{abstract}
\maketitle
\section{Introduction}
\label{intro}
The quantum mechanics of wave phenomena in several low dimensional networks led to 
unprecedented success in understanding the localization phenomena in those model
geometries. Technologically harnessing the advances in lithography techniques, in turn, led to
a scale of scientific impacts that cannot be understated. However, many of the well-studied 
solid state materials are not so ordered.

The suppression of diffusion of wave train in disordered environment has been extensively 
studied for past sixty years, starting with the pioneering work of Anderson~\cite{anderson}.
According to this prescription, random disorder can eventually localize the incoming excitation
depending on the degree of disorder leading to the formation of exponentially localized states.
On the contrary to the canonical case of Anderson localization (AL), wave excitation can
be \textit{locked} in space inside a sub-lattice part of the network.
In the present decade, the physics related to the flat band systems~\cite{suther}
 is on the spotlight and grabs the attention of both theoreticians and experimentalists because these create open platform
for several novel exciting physical phenomena~\cite{lieb}-\cite{chamon}
 carrying the information of quantum physics together with the topological effect.
This type of 
unconventional localization is expected to be found in different quasi-one dimensional or two-dimensional
systems containing one or more dispersionless flat bands~\cite{mati}-\cite{bp2}. Over the recent 
years, considerable attention has been paid to different non-trivial geometries for showing 
flat band localization with short range fluctuation.
Those states are named as \textit{compact localized states} (CLS)~\cite{dias}-\cite{flachreview} 
where wave function amplitude strictly
vanishes beyond a finite sub-part of the entire system.
One such \textit{characteristic cluster} is \textit{effectively} isolated from the others by some selective
vertices where there are no permission for the electron to stay.
These \textit{self-localized} states are caused by phase cancellation
induced by destructive quantum interference in the presence of local spatial
symmetries.
Thus the standing-wave like amplitude profile makes the mobility of the particle as
 extremely low and hence its effective mass becomes diverging one such that the kinetic contribution
 of the wave becomes quenched. This severely affects 
 the transport of the quantum mechanical wave packet which results in
some unconventional consequence like the inverse Anderson transition~\cite{goda}.
The immediate consequence of this is the 
 dispersionless momentum insensitive contribution to the $E-k$ plot.
 This construction is essentially a part of a wider class of phenomena known as Aharonov-Bohm caging~\cite{vidal}  that has been experimentally observed in many systems~\cite{sebabrata}.

 The immediate important application of CLS lies in the fact of information transmission~\cite{mejia}-\cite{xia}
  avoiding any cross talk~\cite{mejia2}
  between them and this directly
 stems from the compactness of such modes.
 For flat band models with finite range hoppings, the compact localized states having a macroscopic
  number of degeneracy are expected not to show any evolution dynamics, i.e., because of cluster-localization,
  they do not spread out spatially during evolution.
 Hence, from the experimental point of view, it is much less challenging to excite a CLS
 than a regular resonant mode.
 Another remarkable thing is that flat band states eventually enable the appearance of distinct
 bound states within a scattering continuum~\cite{von,herr} caused by
 otherwise periodic system.
 
 Significant attention is presently devoted to this field of study due to its experimental
 realization~\cite{seba,vicen} present in diverse topics of condensed matter physics beyond its original
 theoretical proposition.
 The specific topological character of the lattice has a strong influence on the overall spectrum and in certain
cases it can induce exotic spectral features through a set of macroscopically degenerate dispersionless flat
bands. Low-dimensional networks within the tight-binding formalism provide ample example for that.
Several model networks have been studied in detail
in the context of ferromagnetism originating from the flat band~\cite{lieb}-\cite{tasaki2}.
 The nearly flat band states having
nonzero Chern number are found to unravel an interesting physics related to the fractional quantum Hall
effect~\cite{kapit}-\cite{chamon}.
Also by virtue of extremely
low group velocity of the particle, the divergence of the local 
density of eigenstates corresponding to the FB
state has an important impact on the field of thermoelectric devices to enhance their thermopower~\cite{kondo}-\cite{aritra2}.
Since the kinetic part is completely quenched, the dominant nature of the interaction makes the
FB an ideal platform of investigating the many-body physics regarding strongly correlated electron systems
that also includes ferromagnetism.
 
 In this communication, we unravel an analytically exact scheme to work out dispersionless
 compact localized modes in a quasi-one dimensional geometrically frustrated tailor-made system.
 The
motivational aspects behind the present analysis is two-fold.
 We shall highlight the issue of compact localization induced by destructive phase decoherence
 in this overall periodic square-kagom\'{e} ladder network in the first part of the present work.
 However, in the subsequent part of this discussion we look into the impact of applying uniform magnetic perturbation
 on such states. 
 Perturbation in general is expected to disturb the compactness
 of the self-localized modes.
Also, presence of external agent of energy that flat band mode may exhibit complicated dynamics like Bloch
oscillations associated with Landau-Zener tunneling~\cite{cai,sergej}. Sometimes, those degenerate states may
appear to be completely robust against the application of perturbation as in the case of lieb lattice (flat
band at $E = 0$). But the stability of the state solely depends on the lattice geometry. The insensitivity of
flat band mode in respect of the perturbation is recently established experimentally in case of photonic
rhombic lattice~\cite{seba2}.
 In our article, quite interestingly, the application of external flux invites a non-trivial
 flux tunability of the positions of the FB states. A periodic network generally possesses 
 resonant band structure populated by extended kind of Bloch-like eigenfunctions. But owing to
 the phase cancellation, the existence of self-localized modes and its positional engineering using 
 an external parameter makes this problem more interesting as well as challenging also from
 the experimental perspective. The present model thus
gives an achievable
challenging scenario to the experimentalists in this time-zone of significantly advanced
nanotechnology and lithography techniques.
 Needless to say that the electronic scenario is likely
 to be equivalent case for photonic modes as well~\cite{atanu4}. If this happens to be true, 
 then the proximity of
 edge localized non-dispersive states to the completely transparent states may lead to a 
 new exciting physical situation where a comprehensive manipulation can be done over a possible
 transition (\textit{on-off}) between the bound and resonant states as a function of the 
 energy (or frequency in the analogous context of optics) of the excitation.
 
 Moreover, our anticipation is that, a modulation in the
strength of the magnetic perturbation may lead to an engineering of the curvature
 of the dispersion plot. One can therefore obtain a
comprehensive tuning of the mobility and the effective mass tensor
of the electron by virtue of an external agency.
Another meaningful point of interest in this study is to describe
the distribution of the wave function amplitudes of these compact localized states in
real space.
We prescribe a simple method to circumvent
this problem by using real space renormalization group technique.
By virtue of this, we demonstrate here a simple elaborative analytical method to find out the
compact localized eigenstates which are confined on some
characteristic clusters of the entire network. The non-dispersive
behavior of those states is explicitly computed.
It is needless to say that this scheme can be easily extended to any kind 
of low dimensional model structures, even in case of quasiperiodic or
fractal type of geometries. Our analysis thus provides a generalized scheme 
to detect such localized modes exactly.

Before we conclude this discussion, it becomes worth mentioning
that the strength of our analytical attempt is that
recently the well studied flat band lattice structures can be
fabricated~\cite{seba2} in a sophisticated manner by the experimentalists 
to visualize the theoretical outcomes in reality.
 This makes the topic more interesting 
even in the context of slow light
propagation which immediately invites the concept of 
``spatial compression of light energy"~\cite{baba} and its immediate important practical implications.
This present
model, to our mind, can therefore be tested with an appropriately
fabricated model network.

In what follows we demonstrate our findings. In Sec.~\ref{model} we
discuss the model system using the tight-binding Hamiltonian. Section~\ref{density}
contains a rigorous analysis regarding the general spectral character corroborated by numerical 
evaluation of density of modes, transport,
existence of dispersionless states, and the dispersion
relationship. Also in this section we discuss a mathematical scheme to discern the FB modes.
In the next Sec.~\ref{flux} 
we provide a study on application of magnetic flux along with the relevant persistent current
analysis. In the Sec.~\ref{expt} we shall highlight some points regarding the possible scope for experiment.
Sec.~\ref{slow} contains discussion related to the analogous optical case.
Ultimately, in Sec.~\ref{conclu} we draw a summary of our work.

\section{Model and Hamiltonian}
\label{model}
\begin{figure}[ht]
\centering
\includegraphics[clip,width=8.5 cm,angle=0]{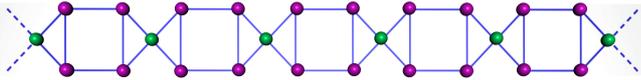}
\caption{(Color online) A portion of a two-legged square-kagom\'{e} ladder network with endless axial extent. 
Violet colored sites ($C_3$) have the number of nearest neighbor equal to three and green
colored quantum dots ($C_4$) have the coordination number four.}  
\label{lattice1}
\end{figure}

In this section we present a outline of our methodology to discern the non-dispersive flat band
states and justification of the localization property of such non-diffracting macroscopically degenerate
states for a geometrically frustrated quasi-one dimensional periodic lattice structure. For this we first
start with the introduction of the model system which is schematically shown in Fig.~\ref{lattice1}. 
The tight-binding Hamiltonian written in the Wannier basis for the non-interacting electrons, that describes 
our system, is given by
\begin{equation}
H  = \sum_{j} |u_{j} \rangle \epsilon_j \langle u_{j}| 
+\sum_{\langle jk \rangle} \left[ |u_{k} \rangle \mathcal{T}_{jk} \langle u_{j}| 
+ h.c. \right] 
\label{hamiltonian}
\end{equation}
The first factor $\epsilon_j$ denotes the potential information at the respective quantum dot location $j$
while the second part $\mathcal{T}_{jk}$ in the Hamiltonian, called \textit{overlap parameter} or hopping integral,
carries the kinetic signature.
It is to be noted that we restrict ourselves to consider the nearest neighbor interaction term only.
Also depending on the local environment i.e., the number of nearest neighboring sites we thus have two 
different kinds of atomic sites present in the square-kagom\'{e} structure.
One with coordination number equal to four and is therefore designated as $C_4$ sites and the another type
having the number of nearest neighbor equal to three and hence is called $C_3$ sites. Therefore the 
on-site energies can take two values $\epsilon_3$ or $\epsilon_4$ based on the coordination number.
The hopping integral is taken as uniform throughout the lattice. Also since we are interested in the 
topological aspects the numerical values of both the two types of on-site terms are uniformly assigned
as zero throughout the numerical calculation and that of hopping is taken as unity without any loss of
generality.

To start the analysis we first start with the alternative discretized form of Schr\"{o}dinger's 
equation i.e., \textit{difference equation}, viz.,
\begin{equation}
( E - \epsilon_j ) \psi_j = \sum_{k} t_{jk} \psi_{k}
\label{difference}
\end{equation}
where the summation on the R.H.S. of Eq.~\eqref{difference} includes all the contributions 
coming due to nearest neighboring sites. 
Momentum insensitive flat bands (FB) for tight-binding quasi-one dimensional square-kagom\'{e}
lattice occurs due to destructive kind of quantum interference leading to a macroscopic 
number of \textit{compact localized eigenstates} (CLS) which have strictly decaying amplitude
beyond a finite sublattice sites of the entire network. Following the real space renormalization
group scheme we describe an analytical approach to compute exactly the non-dispersive states
that are eventually localized even though there is no disorder.

The difference equation allows us to ``decimate out" an appropriate subset of atomic sites
\begin{figure}[ht]
\centering
\includegraphics[clip,width=8.5 cm,angle=0]{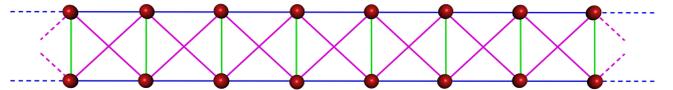}
\caption{(Color online) The \textit{renormalized} two-arm ladder network. The effective ladder
now contains intra-arm connection $\tau$ (blue color), inter-arm connection $\gamma$ 
(green color)
and the second neighbor
hopping $\xi$ (pink color). 
The upper and lower arms of the ladder are respectively designated by $A$ and $B$.
All the sites have the same on-site potentials equal to $\tilde{\epsilon}$.}  
\label{ladder}
\end{figure}
in terms of the amplitudes of the sustained nodal points. This procedure prepares a 
\textit{renormalized} version of two-legged ladder network with \textit{effective} energy
dependent parameters. All the atomic sites in the both `up' and `down' arms of the ladder
geometry have the same on-site potential $\tilde{\epsilon}$. There exists a renormalized overlap
parameter along the arm of the ladder which is $\tau$ and the inter-arm vertical
connection turns out to be $\gamma$. This typical decimation generates a second neighbor
hopping integral as illustrated in the Fig.~\ref{ladder} and is denoted as $\xi$.
The explicit forms of the decoupled parameters can be written in terms of the parameters of the
original network as 
\begin{eqnarray}
\tilde{\epsilon} & = & \tilde{\epsilon_3} + \frac{t^2 (E - \tilde{\epsilon_3})}{\Delta} + \frac{2 t_1^{2} (E -\tilde{\epsilon_3} + t_2)}{\Delta} \nonumber \\
\gamma & = & t_2 + \frac{t^2 t_2}{\Delta} + \frac{2 t_1^{2} (E -\tilde{\epsilon_3} + t_2)}{\Delta} \nonumber \\
\tau & = & \frac{t t_1 (E -\tilde{\epsilon_3} + t_2)}{\Delta} \nonumber \\
\xi & = & \frac{t t_1 (E -\tilde{\epsilon_3} + t_2)}{\Delta}
\label{para}
\end{eqnarray}
In the above Eq.~\eqref{para}, $\tilde{\epsilon_3} = \epsilon_3 + \frac{t^2}{(E - \epsilon_4)}$, $t_1 = \frac{t^2}{(E - \epsilon_4)}$, $t_2 = t + \frac{t^2}{(E - \epsilon_4)}$ and $\Delta = (E- \tilde{\epsilon_3})^2 - t_2^{2}$.

With all these renormalized parameters one can easily write down the difference equation
for the effectively transformed ladder network as follows,
The equation can be written as,
\begin{widetext}
\begin{eqnarray}
(E - \tilde{\epsilon}) \psi_{n,A} & = & \tau (\psi_{n+1,A} + \psi_{n-1,A}) + 
\gamma \psi_{n,B} + \xi (\psi_{n+1,B} + \psi_{n-1,B}) \nonumber \\
(E - \tilde{\epsilon}) \psi_{n,B} & = & \tau (\psi_{n+1,B} + \psi_{n-1,B}) + 
\gamma \psi_{n,A} + \xi (\psi_{n+1,A} + \psi_{n-1,A}) 
\label{difflad}
\end{eqnarray}
\end{widetext}
The above equation for the coupled system can be easily cast in a compact matrix form using the potential and hopping matrices in the form,
\begin{widetext}
\begin{eqnarray}
\left [
E \left( \begin{array}{cccc}
1 & 0 \\
0 & 1
\end{array}
\right ) - 
\left( \begin{array}{cccc}
\tilde\epsilon & \gamma \\
\gamma & \tilde\epsilon
\end{array}
\right)
\right ]
\left ( \begin{array}{c}
\psi_{n,A} \\
\psi_{n,B}  
\end{array} \right )
& = & 
\left( \begin{array}{cccc}
\tau & \xi \\ 
\xi & \tau 
\end{array} 
\right)
\left ( \begin{array}{c}
\psi_{n+1,A} \\
\psi_{n+1,B}  
\end{array} \right )
+
\left( \begin{array}{cccc}
\tau & \xi \\                                                  
\xi & \tau
\end{array}
\right)
\left ( \begin{array}{c}
\psi_{n-1,A} \\
\psi_{n-1,B}
\end{array} \right )
\label{eqladder}
\end{eqnarray}
\end{widetext}
It is very straightforward to check that the forms of the potential matrix
and the hopping matrix are such that they will commute with each other and hence 
can be simultaneously diagonalized using a similarity transform. 
We are now led with a new \textit{decoupled basis} which is related to the old
basis via the relation $\phi_n = \mathbf{M}^{-1} \psi_{n}$. 
The matrix $\mathbf{M}$ diagonalizes both the potential and hopping matrices.
By virtue of this decoupling
technique from the above coupled difference equation we can easily obtain a set of 
linearly independent two decoupled difference equations for the two arms $A$ and $B$
in the transformed basis.
The equations are respectively given by,
\begin{eqnarray}
\left[ E - (\tilde{\epsilon} + \gamma) \right] \phi_{n,A} & = & (\tau + \xi) (\phi_{n+1,A} + \phi_{n-1,A}) \nonumber \\
\left[ E - (\tilde{\epsilon} - \gamma) \right] \phi_{n,B} & = & (\tau - \xi) (\phi_{n+1,B} + \phi_{n-1,B}) 
\label{decoup1}
\end{eqnarray}

It is important to point out that there is no coupling term involving the inter-arm
connection present in any of the above decoupled equations.
These two equations represent two \textit{isolated} one dimensional chains of atoms, each of which 
has identical on-site parameters.
Since the new basis state is related to the old one through a similarity transformation, it is
needless to say that the newly created pseudo-particle state can be obtained from the linear 
combination of the old Wannier orbitals. Hence we can safely comment that each of the above decoupled
equations will produce their own spectral landscape and one can obtain the spectral behavior of the 
entire system from the convolution of those two spectra and physics-wise we will not loose any 
spectral information indeed.

We now proceed to explore these decoupled set of equations to unravel the non-dispersive
bound states, if any, and the other expected resonant modes for this square-kagom\'{e} network.

\section{Spectral analysis}
\label{density}
\subsection{General spectral character}

Before looking for the dispersionless flat electronic bands, it is appreciable to
have a general idea of the characteristic spectrum that gives the complete 
information about the allowed eigenstates of such system. It is seen from
Eq.~\eqref{decoup1} that in each of the decoupled equations, the on-site
and overlap parameters are energy dependent. 
As expected, the complete spectral behavior solely depends on the above two equations.

\begin{figure}[ht]
\centering
\includegraphics[clip,width=5.5 cm,angle=0]{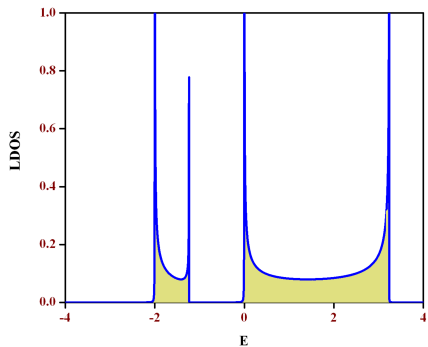}
\caption{(Color online) Local electronic density of states profile as a function of energy. We have set
the numerical values of all the on-site potentials equal to zero and hopping $t = 1$.}  
\label{dos1}
\end{figure}

However, we have employed RSRG decimation scheme on the quasi-one dimensional 
square-kagom\'{e} network. It is now quite straightforward to compute the local
density of states following the standard Green's function formalism and thereby using
the well-known expression
\begin{equation}
\rho (E) = -\left(\frac{1}{N \pi}\right)~Im\left[\bf{Tr}~\mathcal{G}(E)\right]  
\label{green1}
\end{equation}
where, $\mathcal{G}(E) = \left[E + I \delta - \mathcal{H}\right]^{-1}$ and $\delta$ is significantly small quantity which is added to the energy of the incoming projectile. 
`Tr' is the trace of the Green's function.
 We refrain ourselves from incorporating the detailed mathematical
calculations exploited to get the final result just to save space. Here we provide a
typical density of states pattern (Fig.~\ref{dos1}) as a function of the energy of the incoming particle
for the underlying quasi-one dimensional lattice.
The features of flat band modes can be viewed from the portrait of density of states (DOS) in the tight-binding prescription. One can easily endorse that the candidates of the flat band states are essentially localized and condensed in that regime after loosing their mobility.
The Fig.\ref{dos1} also depicts the presence of sharply edge localized spiky states at that flat band energies and these sketch assures the existence of flat band eigenstates in the model system.

The Fig.~\ref{dos1} shows two resonant subbands populated by extended kind of
eigenfunctions. The transparent character of the eigenstates residing on the
absolutely continuous subbands can be checked easily if we look into the flow of hopping integral.
It is seen that the overlap parameter shows oscillatory behavior under successive recursive steps.
This means that the corresponding single particle wave function has never-ending overlap
which is a typical signature of the corresponding states being extended.

\subsection{Analytical construction of compact localized state}
\label{compact}

The physical reason behind the group velocity of the wave packet
 corresponding to flat band mode across a full
 lattice being practically zero is the local geometric phase
 cancellation at the connector nodes. This brings forward the ``trapping" of energy which originates due to destructive interference prevailed by the multiple quantum dots.
 The non-zero amplitudes may be confined within the \textit{characteristic trapping cell} and one such finite cluster
 is `isolated' from the other cluster by a special set of sites on which the wave function amplitude is
 zero.
 This immediately implies the immobility of the wave train as it eventually gets trapped inside the 
 finite cluster of the sites having \textit{non-zero} amplitudes.
 Thus the extremely low or even zero group velocity of the
particle corresponding to such self-localized states eventually
 contributes to the non-dispersive signature. Due
to macroscopic degeneracy, any superposition of those
states is completely static, displaying no evolution dynamics. 

An
 interesting point about
 the flat band is that the dispersion curve is essentially
 flat in curvature in the sense that the single particle energy state is independent of the momentum of the
  particle. This brings a concept of divergent effective mass
  $m^{*} =\hbar^2 / \frac{d^2 E}{d k^2}$
 of the particle leading to practically zero mobility. As a consequence of this, the particle 
 behaves like a \textit{super-heavy} one and the immediate immobility 
 of course severely affects the transport of the quantum mechanical wave packet. 
In other words, the wave packet having such specific FB energy, once locked within the finite size characteristic trapping cell will not get permission to move through the infinite lattice. 
 Since the kinetic
 information is completely quenched, the dominant nature of the interaction makes the FB an apt platform of
studying strongly correlated many-body physics including ferromagnetism.
Needless to
 say that the singular behavior in the density of states profile caused by diverging effective mass
will produce anomalous behavior in the physical
 properties, transport and optical responses. 
 Now we shall raise a valid question that is is possible to extract such bound state eigenvalues for square- kagom\'{e} ladder network?

It is easy to check that if we set $E = \epsilon-2t$ or $E = \epsilon$ (if we set$\epsilon = 0$ and $t = 1$ then $E = -2$ and $0$ respectively), then for each of these two, a consistent solution to the difference equation can be worked out for which the amplitudes will be concentrated within the square loop. Each square loop is decoupled from the adjacent square by the $C_4$ vertices where occupation probability is zero due to zero wave function 
amplitude. This brings essence of cluster localization. 
Since the amplitudes vanish at some selective nodal sites it is possible to \textit{confine} the dynamics of the electron along the boundary of the square loop. This eigenstate so constructed resembles the concept of a molecular state~\cite{egg} that is localized in space partly by the phase decoherence appearing due to destructive quantum interference and partly by the physical boundary formed by the nodes with zero amplitudes. The wave function describing the CLS energy does not have any overlap with that described on the neighboring cell.

The flavor of localization can be understood easily if we study the decoupled set of equations.
\begin{figure}[ht]
\centering
\includegraphics[clip,width=6 cm,angle=0]{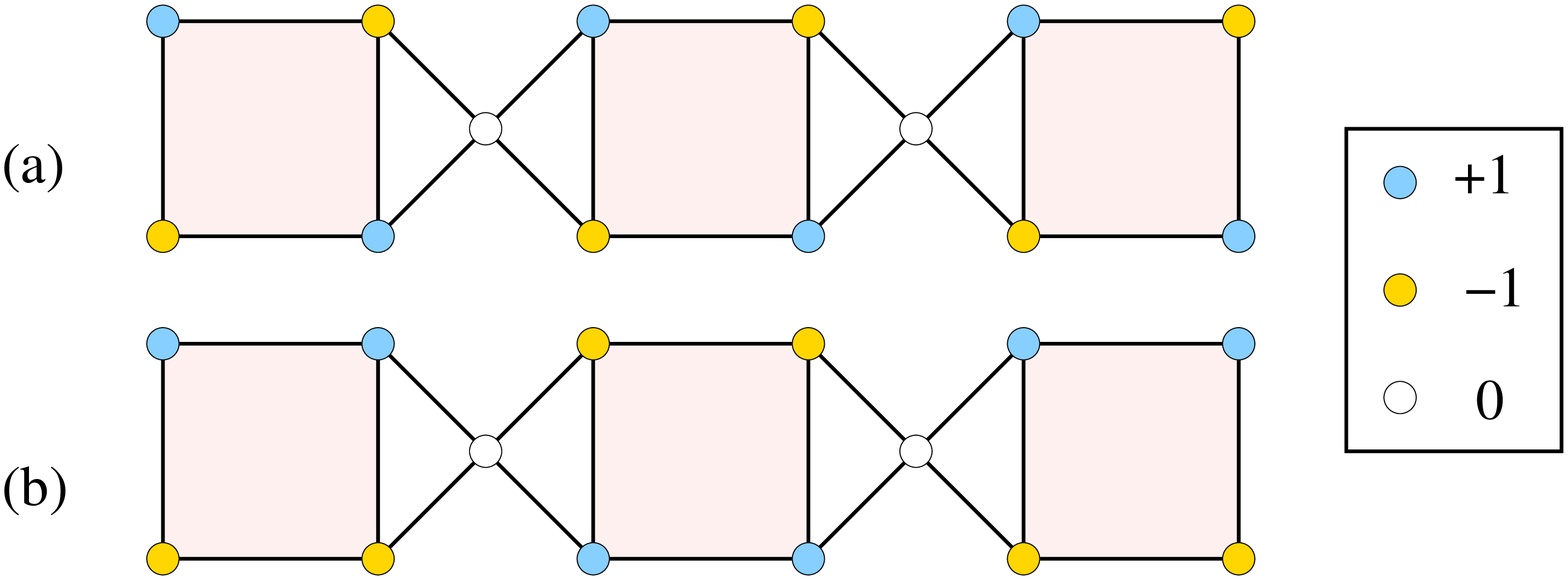}
\caption{(Color online) The amplitude profile showing the essence of cluster localization
for (a) $E = -2$ and (b) $E = 0$.
We have set
the numerical values of all the on-site potentials equal to zero and hopping $t = 1$.}  
\label{amplitude1}
\end{figure}
Since the forms of the inter-arm connection $\tau$ and second neighbor hopping $\xi$ are same, we
can say from Eq.~\eqref{decoup1} that the hopping part of the second equation becomes zero.
Hence, it implies that the envelope of the wave function
corresponding single particle eigenstate essentially 
 has no 
overlap with that of the nearest neighboring sites.
Now if we set all $\epsilon = 0$ and $t = 1$, we get $E = -2, 0$. This equation, speaking in the tight-binding language, now turns out to be a difference equation of a decoupled quantum
dot as the overlap connection vanishes corresponding to such energy. The corresponding single particle eigenstate is essentially compact localized state for which the kinetic part is quenched with 
the strict condition that one has to get non-zero value of density of states at those specific energies. The spatial localization of the bound state is shown in the amplitude profile as cited in the Fig.~\ref{amplitude1}. These two CLS correspond to the two spikes appeared in the two left edges of the DOS spectrum. These are actually edge localized states which have momentum independent contribution to the $E-k$ profile. Due to the zero value of hopping parameter the particle will be locked inside the finite
size cluster as shown pictorially.
\begin{table}[ht]
\caption{Divergence of density of states for the FB mode}
\centering
\begin{tabular}{c c} 
\hline\hline 
Imaginary part of energy ($\eta$) & LDOS ($\rho$) ($E = -2$)\\ [0.5ex]
\hline
$10^{-2}$ & $8.356$ \\
$10^{-3}$ & $80.836$ \\
$10^{-4}$ & $799.754$ \\
$10^{-5}$ & $7970.329$ \\
$10^{-6}$ & $79617.139$ \\
$10^{-7}$ & $795663.632$ \\ 
$10^{-8}$ & $8144999.894$ \\[1ex]
\hline
\end{tabular}
\label{table:diverge}
\end{table}
Also, corresponding to those edge localized states the expected singularity of DOS can be easily
studied as we gradually decrease the imaginary part added to the energy for numerical calculation of
DOS. This is due to the fact that the group velocity vanishes and hence divergence occurs via the following
relationship 
\begin{equation}
\rho \propto \int v_g^{-1} d^3 k.
\label{div}
\end{equation}
 This is tabulated in the adjacent table for the CLS $E=-2$.
Thus a diverging DOS which is a indicative signature of zero mobility confirms the non-dispersive character
of the \textit{self-localized} states as the effective mass tensor becomes also infinitely large.

\subsection{Transmission characteristics} 
\label{tr1}
\begin{figure}[ht]
\centering
\includegraphics[clip,width=8.5 cm,angle=0]{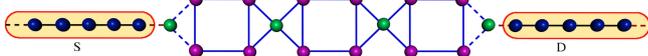}
\caption{(Color online) A finite size square-kagom\'{e} network is clamped in between the
`source' (S) and `drain' (D).}  
\label{lead}
\end{figure}

To corroborate our findings we have also worked out the two-terminal transport characteristics of a finite system. The formalism is quite straightforward and is generally used to compute the nature
of conductivity of any low dimensional network. The central idea is to sandwich the underlying network in between 
a pair of semi-infinite ordered leads, the so called `source' and the `drain' as cited in the Fig.~\ref{lead}.
 
\begin{figure}[ht]
\centering
\includegraphics[clip,width=5.5 cm,angle=0]{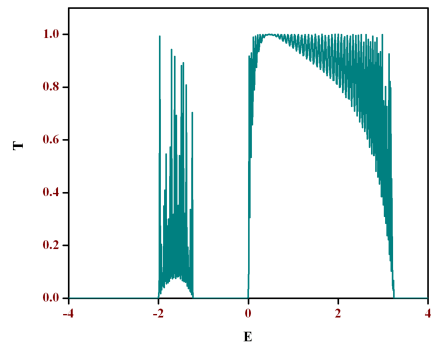}
\caption{(Color online) Two-terminal transport characteristics against energy of the electron.
We have taken
the numerical values of all the on-site potentials equal to zero and $t = 1$.}  
\label{trans1}
\end{figure}
The on-site energies of the quantum dots in the leads are ${\mathcal{U}}_{L}$. 
The overlap parameter is ${\mathcal{T}}_{L}$ in between the 
nearest neighboring atomic sites 
in the leads. The finite sized geometry clamped in between the 
two periodic leads is then recursively renormalized to convert it into an 
effective \textit{dimer} having energy dependent parameters. The transmission coefficient of the 
source-network-drain system can be calculated from the following well-known expression~\cite{liu},
\begin{eqnarray}
& T=\frac{4\sin^{2}qa}{|\mathcal{X}|^{2}+|\mathcal{Y}|^{2}} \\
&\textrm{with,}\quad \mathcal{X}=[(\Sigma_{12}-\Sigma_{21})+(\Sigma_{11}-\Sigma_{22})\cos ka] \nonumber\\
&\textrm{and}\quad \mathcal{Y}=[(\Sigma_{11}+\Sigma_{22})\sin ka]\nonumber
\end{eqnarray}.
Here, $\Sigma_{ij}$ are the dimer-matrix elements and 
$\cos qa=(E-{\mathcal{U}}_{L})/2{\mathcal{T}}_{L}$, 
where $a$ is the uniform lattice constant in the leads and this is taken as unity 
throughout the numerical calculation.

This is seen from the plot that
for the selective regimes of Fermi energy transparent character of the
system is justified by the high transmitting behavior corresponding to
those regions. The extended nature of the eigenstates and the consequent chaotic
oscillation of the flow of hopping integral is the reason of getting high conduction.
Another fact is that the transmission shows considerably low value corresponding
to the flat band energy which is expected.

\subsection{Dispersion relation}
\label{dis1}
It is now becoming a very pertinent question that whether this analytically worked out self-localized states are non-dispersive in nature or not.
Here an important
fact to be mentioned that the compact localized states stated
so far have been `constructed' so as to describe a kind of \textit{self
 localization},
where the amplitudes are distributed over a finite size
cluster of atomic sites, effectively `decoupled' from the other part of
the network.
Keeping this in mind we shall use real space renormalization group scheme 
and decimate out `appropriate subset of vertices' to convert the present structure
into a periodic mono-atomic chain with effective energy dependent on-site and overlap 
parameters. The standard description of the dispersion relation for this monoatomic chain
in the tight-binding language as a function of the effective parameters of the chain is 
$E = \epsilon_{eff} + 2 t_{eff} \cos ka$.
\begin{figure}[ht]
\centering
\includegraphics[clip,width=4.5 cm,angle=0]{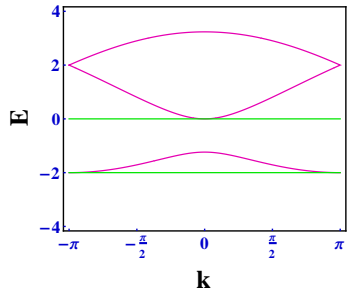}
\caption{(Color online) The electronic dispersion relation for the square-kagom\'{e} ladder
geometry showing the momentum independent contributions at $E=-2$ and $E=0$.}  
\label{disp1}
\end{figure}
 
As mentioned earlier that throughout our analysis,
we have considered the numerical values of the potentials
of all the sites as zero and that of the overlap parameter equal to unity
Then the above expression on simplification can be rewritten as
\begin{equation}
E (E + 2) \left( 4 - 4 E - 2 E^{2} + E^{3} - 4 \cos ka \right) = 0
\label{dispeq}
\end{equation}
The relation is plotted in the Fig.~\ref{disp1}.
It is clearly seen from the dispersion plot the flat band states occur at $E = 0, -2$.
This discussion presents the confirmation of the dispersionless character of the compact localized states discussed in the previous section.

\section{Effect of uniform magnetic flux}
\label{flux}
\subsection{Energy-flux spectrum}

In this section, we study the consequence of applying a uniform magnetic flux on those
FB states along with the study of general character of the spectrum. Each square plaquette
is threaded by a uniform magnetic perturbation that incidentally breaks the time reversal
symmetry of the overlap connection along the bond of the square, and it is taken into consideration
by incorporating a Peierls' phase factor associated with the hopping parameter, viz.,
$\mathcal{T} \rightarrow \mathcal{T} exp \left[ \frac{i e}{\hbar c} \int_{i}^{j} \bm{A}. d \bm{l} \right]
= \mathcal{T} e^{i \frac{2 \pi \Phi}{4 \Phi_0}}$, where $\Phi_0 = h c/e$ is the fundamental flux quantum.

\begin{figure}[ht]
\centering
\includegraphics[clip,width=6 cm,angle=0]{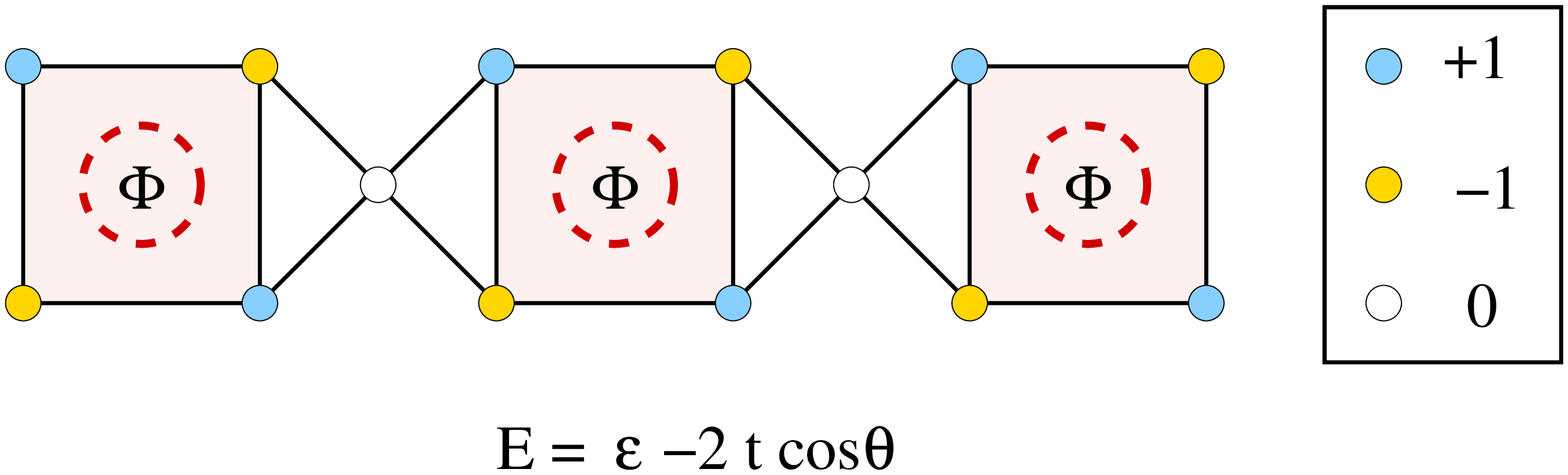}
\caption{(Color online) Amplitude distribution for the FB state $E = \epsilon - 2 t \cos \theta$ in
presence of flux.}  
\label{amplitude2}
\end{figure}

\begin{figure}[ht]
\centering
(a)\includegraphics[clip,width=5cm,angle=0]{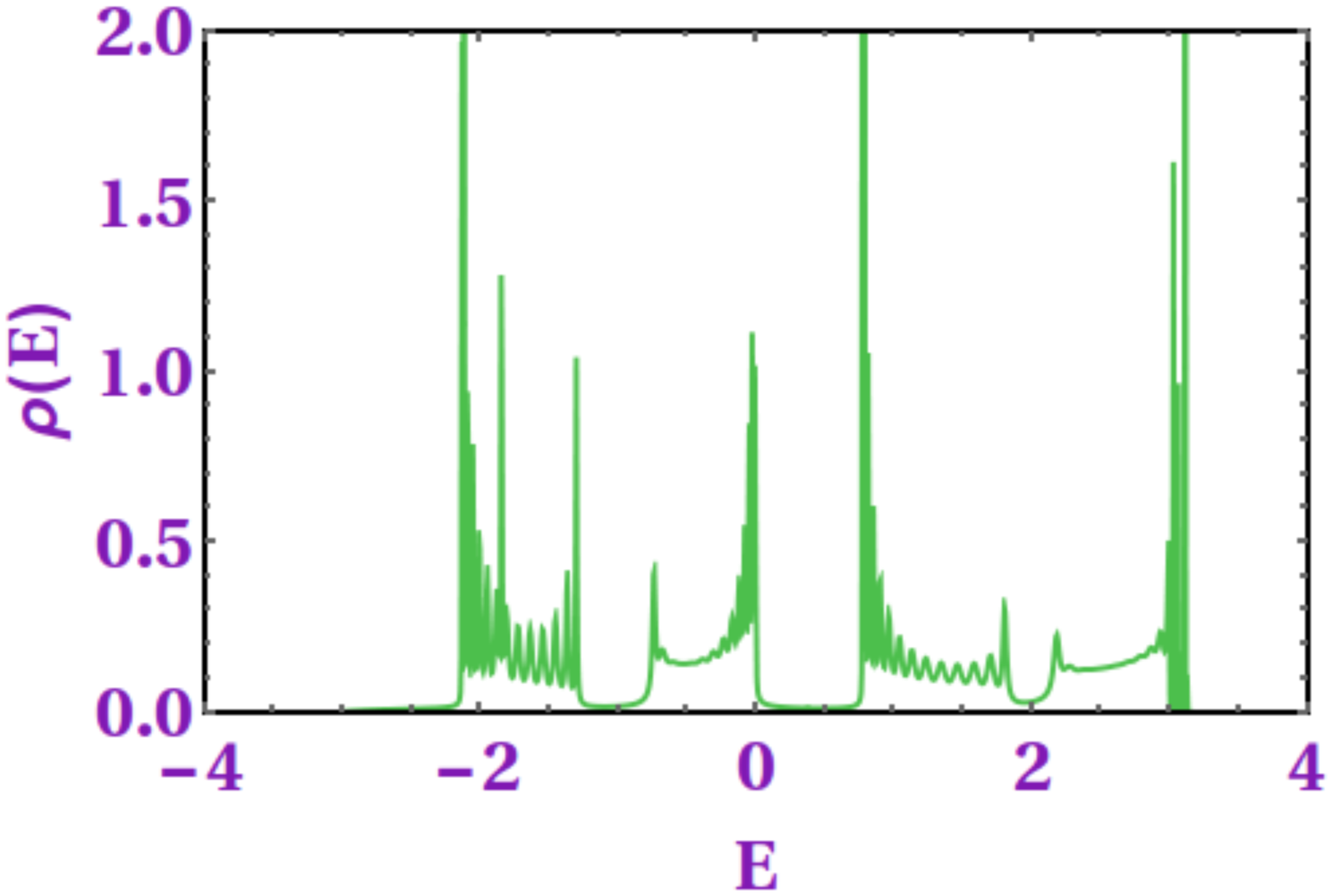}
(b)\includegraphics[clip,width=5cm,angle=0]{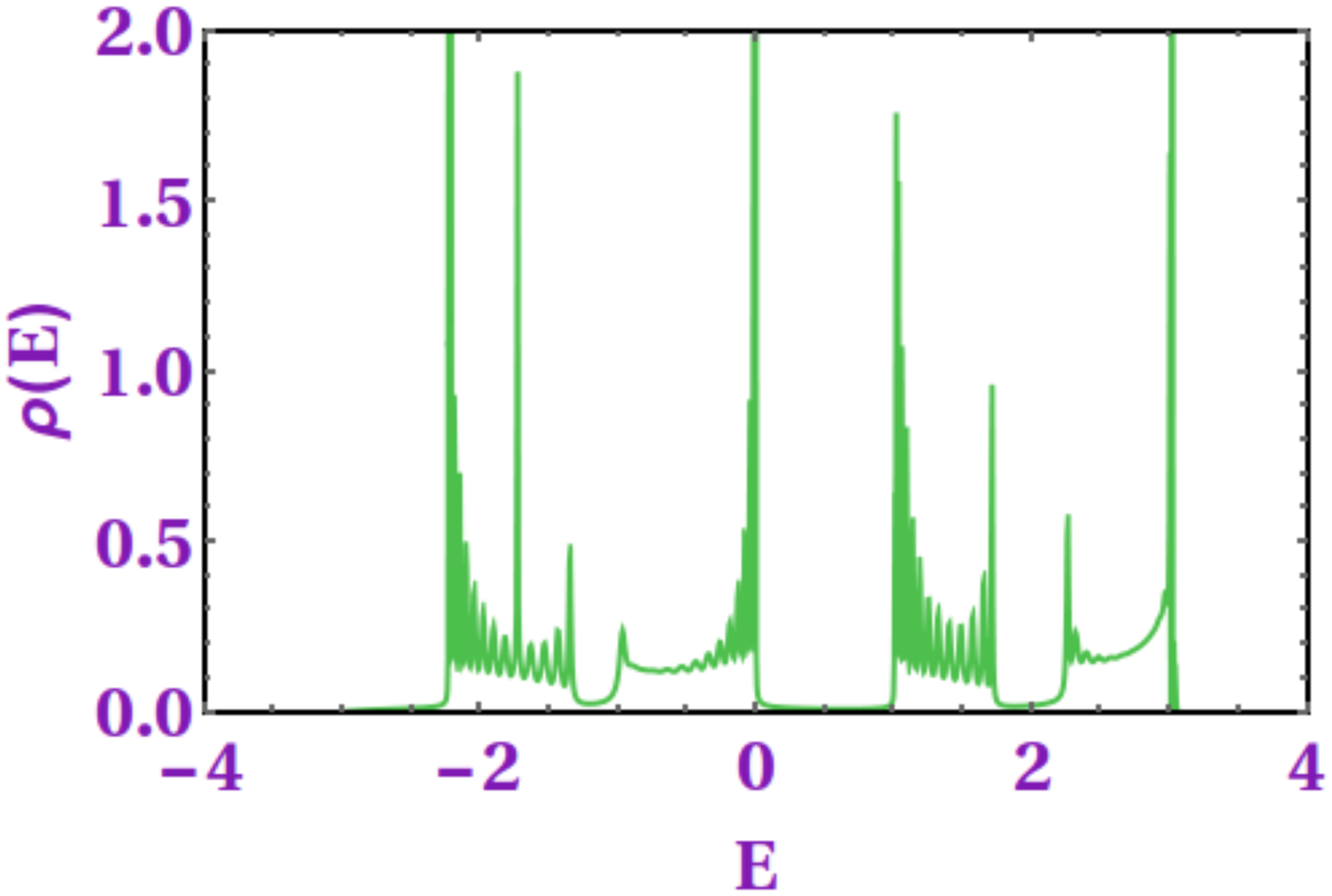}
\caption{(Color online) 
The density of states pattern in presence of flux;(a) $\Phi = \Phi_0 / 4$ and (b) $\Phi = \Phi_0 / 3$.
The momentum independent states are the distinct spikes corresponding to the energies $E = - 1.847$ and 
$E = -1.732$ respectively. We have set $\epsilon = 0$ and $t=1$.}
\label{dosflux}
\end{figure}

\begin{figure}[ht]
\centering
\includegraphics[clip,width=7 cm,angle=0]{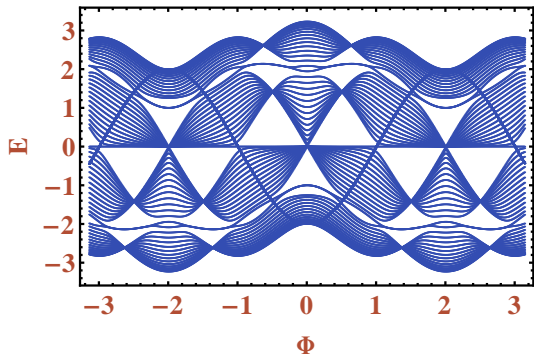}
\caption{(Color online) Allowed eigenspectrum as a function of uniform magnetic flux applied. The
flux dependence of flat band states $E = \epsilon - 2 t \cos \theta$ is clearly visible in the plot.}  
\label{spec}
\end{figure}

The sustainability of the FB states on applying magnetic flux depends on the nature of
quantum interference occurred by the multiple quantum dots. We refer to the Fig.~\ref{amplitude2}.
Following the amplitude distribution shown, using the tight-binding difference equation it is very
trivial to calculate the flux tunable FB state at $E = \epsilon - 2 t \cos \theta$ ($\theta = \frac{2 \pi \Phi}{4 \Phi_0}$ is the typical Aharonov-Bohm phase factor).
Due to the divergence of density of states corresponding to the flat band modes(Eq.\eqref{div}),
 one should observe a distinct spike for such modes. 
We have shown the DOS profile for different values of flux in the Fig.~\ref{dosflux}.
For all values of flux the spectral profile supports the existence of flat band states.
Another interesting point is that we can tune the magnetic flux in a continuous manner to 
engineer the position of such self-localized modes.
This continuity is supported from the allowed eigenspectrum (Fig.~\ref{spec}) also.
Due to the vanishing amplitude at the $C_4$ sites the particle having FB energies will be locked
inside the square plaquette but this time we have now an external parameter
that can be controlled at will. Hence we have the possibility of engineering 
the positions of the flat band states by periodic modulation of magnetic flux applied. 
Position sensitivity to the magnetic flux makes the thing interesting from the 
experimental standpoint.

In Fig.~\ref{spec} we present an energy vs. flux diagram which illustrates the allowed
eigenmodes as a function of the external flux. For an electron moving round a closed
loop, the magnetic flux plays the role same as the wave vector~\cite{gefen}.
 So we demonstrate the $E - \Phi$ profile as an equivalent dispersion relationship. 
 We have taken a finite size square-kagom\'{e} lattice
 of $76$ atomic sites to compute the spectral landscape
 It is clearly seen that the energy eigenvalues of the network are found to form minibands
 as a function of the applied flux $\Phi$. 
 Multiple inter-twined band crossings and a quite dense distribution of eigenmodes 
 produce a \textit{quasi-continuous} $E-\Phi$ band.
 At some points of band crossings, the spectrum shows kind of zero bandgap semiconductor
 like behavior, mimicking Dirac points as observed in graphene, at $\Phi = \pm 2n \Phi_0$ ($n$ being an
 integer).
 It is needless to say that the dense pattern is of course flux periodic with a periodicity equal to
 $4 \Phi_0$. A very minute observation of this energy-flux spectral scenario reveals the creation
 of interesting variants of the Hofstadter butterflies~\cite{butterfly}. Also, it is very pertinent 
 to emphasize that the formation of flat band state at $E = \epsilon - 2 t \cos \theta$ is also supported
 from this allowed eigenvalue distribution.

\subsection{Study of persistent current}
\label{current}
For the sake of completeness of the detailed spectral analysis in presence of
uniform magnetic flux for this quasi-one
dimensional system, we present a study of the behavior of persistent current~\cite{gefen} against the
applied perturbation.
The current is given by
\begin{equation}
I(\Phi) = - k \frac{\partial E}{\partial \Phi}
\label{persis}
\end{equation}
\begin{figure}[ht]
\centering
\includegraphics[clip,width=5 cm,angle=0]{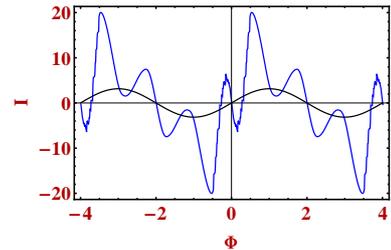}
\caption{(Color online) Study of persistent current for the square-kagom\'{e} network.
The black colored curve shows the contribution from the flux dependent CLS. The
degradation of amplitude of the response 
corresponding to the CLS in comparison with that for the states other than CLS (blue curve)
is apparent from the plot.}  
\label{cur}
\end{figure}
If we have the energy eigenvalues as a function of magnetic flux $\Phi$, the
persistent current for individual energy eigenstates can be easily evaluated using the Eq.~\eqref{cur}. 
The fact is simply
we have to take the slope of the $E-\Phi$ diagram i.e.,
just the first order derivative of energy with respect
to flux $\Phi$.
The result is plotted in the Fig.~\ref{cur} ($\epsilon = 0$, $t = 1$). It shows an oscillating nature of the current (response of the system)
against the applied flux as expected.
Also the inherent periodicity of $4\Phi_0$ of the spectral landscape is clearly reflected in
the variation of current. The magnitude of the current is quite high and this is quite obvious
since a perfectly ordered system in general provides Bloch-like extended states which
carry finite current.
Where as for the phase cancellation the compact localized states do not contribute to
the significant response of the network. Hence the amplitude of the persistent current
will diminish as shown in the Fig.~\ref{cur}.
\section{Possible experimental proposition}
\label{expt}

It is now well-known fact that the localization is very common phenomenon not only in the 
diverse fields of condensed matter physics but also is significantly relevant in the context of
classical waves as well, including light. The journey was initiated by Yablonovitch~\cite{yablo,yablo2} along
with the proposals from John~\cite{john} and Pendry and MacKinon~\cite{pendry} related to the possibility of
experimental observation of Anderson localization. Inspired by all these pioneering works it is
now important to move our focus on the possibility of experimental realization, if possible, of photonic
flat band modes in such quasi-one dimensional structure.
In the present advancement of nanotechnology and lithography techniques, several
authentic examples regarding the experimental aspects are present in the literature.
The most recent experiments by S. Mukherjee \textit{et al.}~\cite{seba,seba2} and 
also by the other authors~\cite{hu,zong}
 are also remarkable milestone
 regarding the photonic localization of wave train in photonic lattices formed by 
laser-induced single-mode wave guides. All these works prompt us to work out the analytical scheme
of extracting such self-localized modes for such model networks. The present study thus 
throws an achievable challenge to the experimentalists to test our analytical result with
suitable fabrication technique.
The wave propagation in such single-mode optical wave guides can be adjusted by
femtosecond laser-writing method as well as the optical induction technique. 
The overlap parameter can be manipulated by the dielectric properties of the core material.
This provides a scope for direct observation of diffraction free FB states.
Also, a synthetic gauge field~\cite{longhi1,longhi2}
 can be generated by modulating the propagation constant
to  study the effect of magnetic field in our proposed geometry.

\section{Slow light analogue}
\label{slow}
Light plays a very important as well as
effective role as a messenger in the communication system and optical communication is one of the well-studied aspect in the technological field. Electromagnetically induced transparency (EIT) has become one of the most effective and powerful technique for engineering the quantum states of light fields~\cite{imam}. A light field that carries the classical (or quantum) information can be slowed down or even gathered
 by tuning the Rabi frequency of the control light field~\cite{mair}.
\begin{figure}[ht]
\centering
(a)\includegraphics[clip,width=5cm,angle=0]{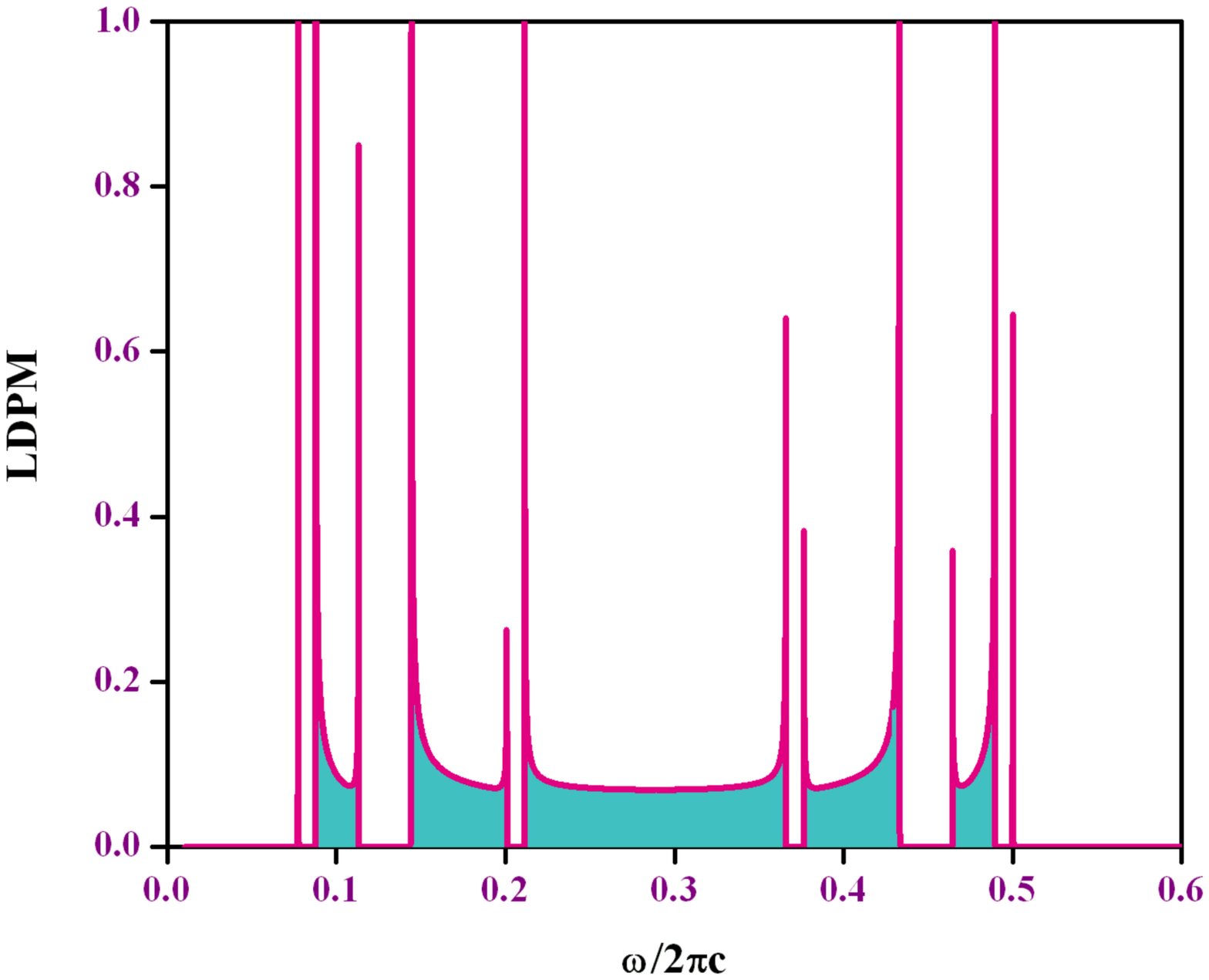}
(b)\includegraphics[clip,width=5cm,angle=0]{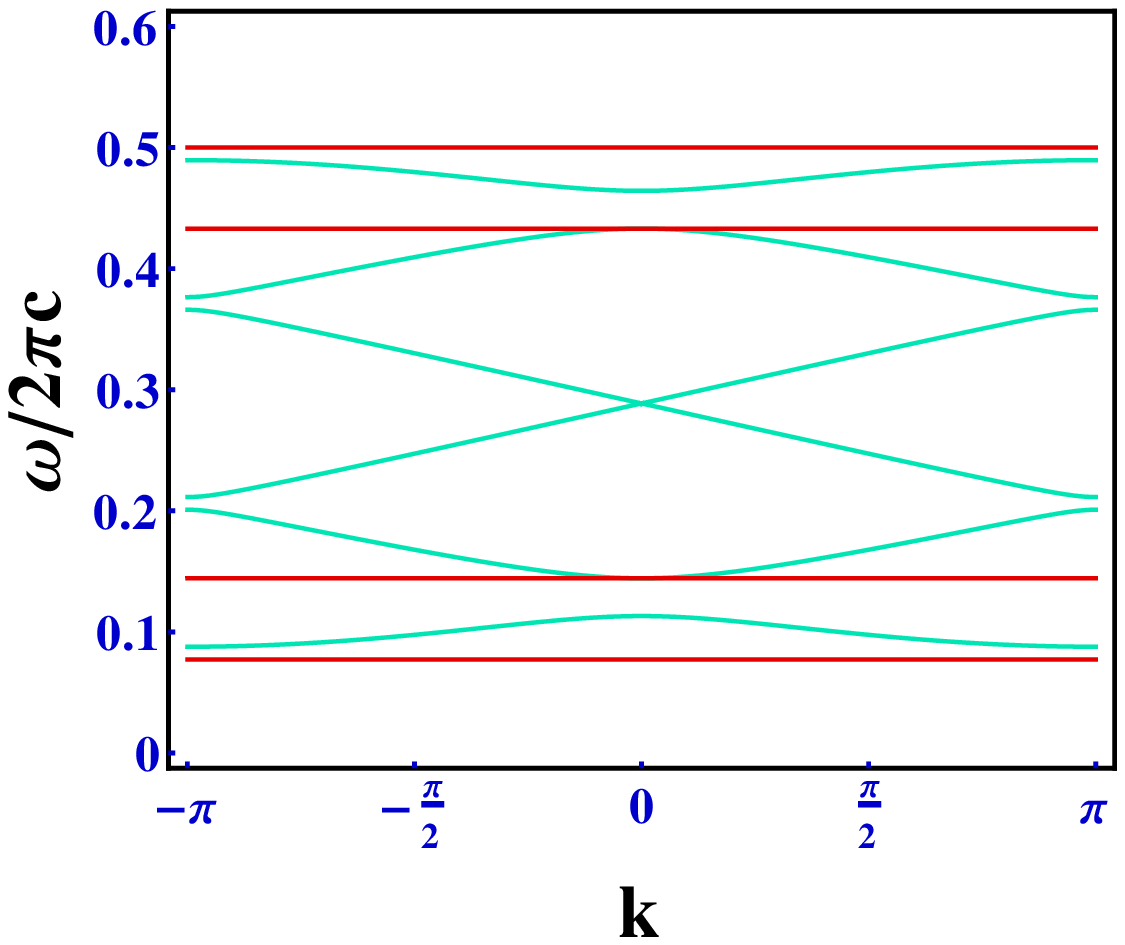}
\caption{(Color online) 
(a) Local density of photonic modes as a function of the frequency of the wave injected 
for the wave guide network arranged in square-kagom\'{e} geometry and
(b) the corresponding optical dispersion relationship.}
\label{opt}
\end{figure}

Slow light with a \textit{remarkably} low group velocity of the wave train is a very promising solution for optical delay line or optical buffering and advanced time-domain optical signal processing. The concept of slow
light enhances linear and nonlinear effects and thus miniaturizes functional photonic devices, due to the
fact that slow light compresses electromagnetic energy in space that eventually increases the light-matter
interaction. This in turn, results in enhanced gain and absorption, phase shift, and non-linearities.

Analogy between the electronic model and the corresponding photonic scenario within the
tight-binding framework is not new. As first introduced by Sheng \textit{et al.}~\cite{wave1,wave2}, the wave propagation through any quasi-one dimensional network can be exactly mapped back onto the corresponding electronic case with the appropriate parametric substitution. 
The equivalence of
 the difference equations with that of an electron moving in
 a similar geometry helps us to study the localization of classical waves. Also, the one-to-one correspondence
  is entirely a mathematical aspects
  and once this is done the recursion equations used for RSRG technique are seen to be
 irrelevant to whether the incoming excitation appears from a quantum background
 or a classical one. Obviously here the propagation of classical wave can be manipulated by selective choice of dielectric parameter. If one sets the dimension
 of the wave guide segment at the very beginning of calculation, then one can
 easily compute the \textit{exact} frequency of injected wave train for which the self-localization occurs.

Following the tight-binding analogy we have also presented (Fig.~\ref{opt}) the 
photonic dispersion relation and local density of photonic modes as a function of the
frequency of the injected photon. The appearance
of slow optical modes (within $0 < \omega/2 \pi c < 1/2$) triggered by quantum interference
is clear in both the plots.

\section{Summary and outlook}
\label{conclu}

In conclusion, we have unraveled an analytical scheme showing the methodology to obtain the eigenvalues corresponding to the dispersion-free degenerate compact localized states of a periodic quasi-one dimensional two-arm square-kagom\'{e} ladder network of endless axial span. 
The flat band states are reinforced by the meticulous numerical workout of the density of eigenstates along
with the distribution of amplitudes in our toy model.
The observation of flat band states for the spinless single particle in the experimental laboratories is recently successfully achieved with the aid of immense development in the technology. There is already an embark of the proposal of flat band modes in the spin hyperspace for spin one ultracold atoms in the physics community. In future one can examine the robustness of the flatness of the energies in the perturbation named as spin-orbit coupling.
Also an applied magnetic flux is shown to be decisive external agency
to tune the position as well as the curvature or the group velocity for such flat band modes. The periodic variation of such FB states with flux leads to an interesting modulation of the compactness of localization that may inspire the experimentalists. The non-dispersive behavior corresponding to
 the FB modes are justified from the energy-momentum dispersion profile and also due to the divergent behavior of the DOS at such specific energies. The two-channel transport mechanism corroborates our findings. The magnetic perturbation induced quasi-continuous subbands have been shown elaborately. 

This scheme is easily extendable to photonic, plasmonic or magnonic excitation for any kind of quasi-one dimensional geometries and the FB states can be exactly evaluated. This may open up new challenges to the experimentalists to control the creation and positioning of the dispersionless energy bands in such artificial lattices. A possible application thus may be proposed in the field of device technology.

\begin{acknowledgments}
A. M. is grateful to DST, INDIA for providing the financial support provided through
 INSPIRE fellowship [IF160437].
Both the authors are thankful for the stimulating
discussions regarding the results with Prof. Arunava Chakrabarti.
\end{acknowledgments} 

\end{document}